\documentstyle[preprint,aps,tighten]{revtex}
\begin{document}

\preprint{IITAP-99-012 \,\,\,  
INR-1029/99 \,\,\,
hep-th/9910157}

\draft

\title{\bf Canonical Formulation of the Light-Front Gluodynamics \\
 and Quantization of the Non-Abelian Plane Waves} 

\author{\large  Victor T. Kim${}^{\dagger \&}$, 
Victor A. Matveev${}^{\ddagger}$, 
Grigorii B. Pivovarov${}^{\ddagger \&}$\\ 
and \\
James P. Vary ${}^{\S \&}$ }

\address{ ${}^\dagger$ :
St.Petersburg Nuclear Physics Institute,
Gatchina 188350, Russia  \\
${}^\ddagger$ :
Institute for Nuclear
Research, Moscow 117312, Russia \\
$\S$ : Department of Physics and Astronomy, \\ 
     Iowa State University, Ames, Iowa 50011, USA \\
$\& $ :
International Institute of Theoretical and Applied Physics,\\
     Iowa State University, Ames, Iowa 50011-3022, USA} 

\maketitle
\begin{abstract}

Without a gauge fixing,
canonical variables for the light-front $SU(2)$ gluodynamics  are
determined. The Gauss law is written in terms of the canonical variables. 
The system is qualified as a generalized dynamical system with
first class constraints. Abeliazation is a specific feature of the 
formulation (most of the canonical variables transform nontrivially only
under the action of an Abelian subgroup of the gauge transformations).
At finite volume, a discrete spectrum of the 
light-front Hamiltonian
$P_+$ is obtained in the sector of vanishing $P_-$. 
We obtain, therefore,  a quantized form 
of the classical solutions previously known as 
non-Abelian plane waves. Then, considering the infinite
volume limit, we find that the presence of the mass gap 
depends on the way the infinite 
volume limit is taken, which may suggest the presence
of different ``phases'' of the infinite volume theory.
We also check that the formulation obtained is in accord with the
standard perturbation theory if the latter is taken 
in the covariant gauges.

\vspace*{1cm}
to appear in {\it Physical Review D}
\vspace*{1cm}
\end{abstract}


\newpage
\section{Introduction}
The light-front formulation of relativistic dynamics
has been widely discussed since the work of Dirac \cite{Dirac}.
Its virtue is the presence of a kinematic
semi-positive observable, the momentum component $P_{-}$.
It generates shifts along the longitudinal direction
$x^-=(x^0-x^3)/\sqrt 2$ at a fixed light-front time
$x^+=(x^0+x^3)/\sqrt 2$. As such, it should be quadratic in
the canonical variables even in the presence of an
interaction (such variables are called kinematic).
For a system with no tachyons ($P^2=P_-P_+-P_\perp^2 \geq 0$, 
$P_0\geq 0$), 
$P_-$ is semi-positive, and the subspace annihilated
by $P_-$ contains the translation invariant vacuum state.
In addition, for a system with no massless states 
({\it i.e.} having  a mass gap), 
the above subspace contains only the states of vanishing
four-momentum. Therefore, in the light-front formulation,
finding the vacuum state and the Fock space of such a system are kinematic
problems.

The light-front formulation also induces a specific
kind of intuition believed to be valuable in many
physical situations (for a review, see Refs. \cite{Brodsky1}).
For example, the light-front formulation turns out to be closely
related to the infinite momentum frame limit \cite{Weinberg}, and to
the notion of constituent quarks \cite{Wil}.
We also note that the light-front formulation in a finite volume 
appears in discussions of 
recent developments of M-theory \cite{Susskind}.

Thus, the light-front formulation of gauge theories and, 
in particular, of gluodynamics, is a worthwhile goal.
However, up to the moment, the light-front formulation of the
gauge theories is less well developed, in our opinion, 
than the conventional
equal-time formulation. The reason for the above assertion is that,
to overcome technical difficulties, the light-front
formulation has been confined to fixed
gauges (almost always, to the light-cone gauge 
\cite{Tomb,Franke,Pauli},
or to other gauges \cite{Prz,Brodsky2}). On the other hand,
a general
formulation of a gauge theory (see, {\it e.g.}, Ref. 
\cite{Faddeev}) may better
start with a determination of the canonical variables 
prior to any gauge fixing.
In the case of the equal-time formulation of the gauge theories, 
the canonical  variables are ${\bf E}$ 
and ${\bf A}$. 
With
this accomplished, one is to find the constraints and calculate
the algebra of the Poisson brackets involving the constraints 
and the Hamiltonian. After that, the
system is qualified as
a generalized dynamical
system with first class constraints (the case of the 
equal-time formulation of the 
gauge theories), and a  well-developed
machinery to treat such a system, in particular, by fixing a
gauge and introducing the Faddeev--Popov ghosts, becomes available.

In this paper, we determine the canonical variables of the 
light-front
$SU(2)$ gluodynamics without a gauge fixing. 
The light-front version of the Gauss law is also 
determined, and the system is
qualified as a generalized dynamical system with first class
constraints. Thus, the light-front formulation of a gauge theory 
is established
on a par with the conventional 
equal-time formulation. 

A specific feature of the formulation obtained is that
it has a form of an Abelian gauge theory, because most of
the canonical variables transform nontrivially only under
the action of the Abelian subgroup of the gauge transformations
which leaves the component $A_-$ of the gauge field
invariant. Thus, we obtain a version of the Abelian projection
\cite{'t Hooft}  without a gauge fixing.

Quantization of the dynamical system obtained involves the ambiguity 
of ordering. We fix it in the simplest way, and then check that our choice
leads to the standard Feynman rules in the covariant gauges.

We also  consider a gauge 
invariant reduction of the dynamics to the configurations
of zero $P_-$. To this end, we diagonalize $P_-$, {\it i.e.},
identify the excitations carrying the nonzero quanta of the
longitudinal momentum. Then, the reduction is obtained by
nullifying the canonical variables corresponding to these
quanta.  If there is a mass gap in gluodynamics,
and the light-front formulation is ``correct'' ({\it i.e.}, equivalent
to the equal-time formulation), we expect the light-front
Hamiltonian $P_+$ to be vanishing on the equations of
motion of this reduced dynamics. While the reduced dynamics
is indeed much simpler than the complete one, the vanishing of 
$P_+$ is not evident. 

This brings up another important issue, namely, the dependence of 
the formulation on the infrared regularization. To determine the canonical
variables, we need to introduce a gauge invariant infrared cutoff.
In this paper, we use a compactification on a torus 
imposing periodic boundary conditions on the gauge fields along
the $x^-$ direction. Then, it turns out that the reduced $P_+$ does vanish
on field configurations decaying fast enough at the infinity
of the transverse plane, and is nontrivial
on the configurations of nonzero asymptotics at the transverse infinity.
Not surprisingly, in the latter case, the above reduced dynamics turns out
to be a dynamics of the zero modes 
\cite{Maskawa}, {\it i.e.}, of the fields
with an imposed dependence on the longitudinal and transverse space 
coordinates. The classical solutions of the gauge field equations
obtained in the framework of the reduced dynamics coincide up to
a gauge with the previously known 
``non-Abelian plane wave'' solutions due to Coleman \cite{Col}. 
The invariance of the
quantum theory with respect to certain ``large'' gauge transformations
gives a quantization condition for these non-Abelian plane waves.
Therefore, the spectrum of $P_+$
on the subspace of vanishing $P_-$ turns out in this case to be discrete,
bounded from below, and the quantum of the spectrum
is proportional to $(g^2 L)/V_\perp$, where $V_\perp$ is the volume
of the compactified transverse plane, $g$ is the gauge coupling, 
and $L$ is the length of the compactified direction $x^-$. Note that
this scaling of the quantum of the light-front energy holds for any 
nonzero number of the transverse dimensions,
and the presence of the coupling makes the dimension correct.

We conclude that there is no mass
gap in the finite volume light-front gluodynamics. 
On the other hand, the spectrum of the light-front
gluodynamics at infinite volume is qualitatively
dependent on the way the infinite volume limit is taken.
If $L$ is taken to infinity first, the mass gap can
be generated. If $V_\perp$ is taken to infinity before $L$, 
the resulting theory, if it exists, has no mass gap.
Therefore, this can indicate that the finite volume theory 
contains markers ($V_\perp$ and $L$) for different ``phases'' 
of the infinite volume theory.

The rest of the paper is organized as follows. 
In Section 2, we define the canonical variables for $SU(2)$ gluodynamics
without a gauge fixing, using Faddeev-Jackiw approach to constrained systems 
\cite{Jackiw}.
Then, in Section 3, we present the light-front Gauss law, 
which, in analogy with
the equal-time formulation, generate
the local gauge transformations of the canonical variables.
In Section 4, we write the equation for the zero modes which
is necessary to determine the light-front Hamiltonian.
In Section 5, we discuss the quantization ambiguity, and check that the
simplest prescription for fixing the ambiguity leads to the standard
Feynman rules in the covariant gauges. 
In Section 6, we show that the reduced dynamics at $P_{-} = 0$ leads
to non-Abelian plane waves solutions, and we find their quantum spectrum.
Section 6 contains discussion and conclusion.

\section{Canonical Variables}

We start with the action of the $SU(2)$ gluodynamics:
\begin{equation}
\label{action}
S_{glue}=\int\,dx^+dx^-dx^\perp \Big[\frac{1}{2}F_{+-}^a F_{+-}^a +
                                 F_{-k}^a F_{+k}^a  - 
                   \frac{1}{2} F_{12}^a F_{12}^a  \Big],
\end{equation}
where $x^\pm=(x^0\pm x^3)/\sqrt 2$; $x^\perp=x^{1,2}$;
$F_{\mu\nu}^a=\partial_\mu A^a_\nu - \partial_\nu A^a_\mu
+ g\epsilon^{abc}A^b_\mu A^c_\nu$ is a strength tensor of 
the gauge field $A_\mu^a$; $a,b,c,...$ are 
color indices running from 1 to 3, and $k$ is a
Lorentz transverse index running from 1 to 2.

Our aim is to give a canonical formulation of the system (\ref{action})
with $x^+$ as time.
To this end, we will make a chain of transformations of 
the field variables. Every transformation
will be one-to-one, or will introduce new auxiliary variables expressible
via the initial variables. 
At each step, we will keep track of the form that the terms
of the action with the time derivatives assume. 
Ultimately we will obtain the
canonical form, $\sum_i p_i\dot q_i$ (the overdot denotes the 
time derivative), 
for these terms, and will recognize
$p_i$ and $q_i$ as canonically conjugated variables. 
This way of treatment is in the spirit
of the Faddeev--Jackiw approach to constrained systems
\cite{Jackiw}.

What complicates this program is the way the time derivatives
of $A_k^a$ enter the action (\ref{action}): 
$(D_-A_k)^a\dot{A}_k^a$, where 
$(D_-\Phi)^a=\partial_-\Phi^a+g\epsilon^{abc}A_-^b\Phi^c$ 
is the covariant derivative in the $x^-$ direction. 
A simplification of this term is the reason
to confine the formulation to the light-cone gauge \cite{Franke,Pauli}.
This approach is not available for us, as we explained above.
Instead, to simplify this term, we suggest a transformation
to new variables for $A_k^a$. We will denote them $\tilde{A}_k^a$.
The correspondence between the initial variables $A_k^a$ 
and the new variables $\tilde{A}_k^a$ taken at the same moment of time
is one-to-one and depends on the configuration of ${A}_-^a$ at the
same moment of time. 

Another ingredient of treating the term with the time derivatives of
$A_k^a$ is taken in concord with Refs. \cite{Franke,Pauli}. 
That is, we  compactify the theory
along the $x^-$ direction. Namely, all the fields
are considered to be periodic in $x^-$: 
$A_\mu^a(x^-=-L/2) = A_\mu^a(x^-=L/2)$. In this case, the
spectrum of $D_-$ is discrete, and it becomes evident that
the components of $A_k^a$ nullified by $D_-$ are  nondynamical.
One may neglect this subtlety at the expense of appearance of
infrared divergences in the formulation.

To begin our chain of variable transformations, we 
start with the less problematic terms with the time derivative of
$A_-^a$.
The first term in the square brackets of Eq. (\ref{action}) contains
a square of the time derivative of $A_-^a$. In the case of the 
equal-time formulation, the time derivatives of all space components 
of the gauge field enter the action in this way. In the case under 
consideration, we will treat
the time derivatives of the component $A_-$ in analogy with the
equal-time formulation \cite{Faddeev}. Namely, 
to get an action
linear in time derivatives, we substitute the action (\ref{action})
by an equivalent action:
\begin{equation}
\label{calE}
S_{glue} = \int\,dx^+dx^-dx^\perp\Big[
{\cal E}^a F_{+-}^a
+ F_{-k}^a F_{+k}^a
-\frac{1}{2}\big({\cal E}^a{\cal E}^a+
F_{12}^a F_{12}^a \big)\Big],
\end{equation}
where ${\cal E}^a$ is considered as an independent variable.
To see the equivalence,
take the variation with respect to ${\cal E}^a$ and substitute back in
(\ref{calE}) its extremal value $F_{+-}^a$. Now Eq. (\ref{calE})
is linear in the time derivatives, and the
content of the round bracket gives the light-front
energy yielded by the Noether procedure: $P_+=\int\,
dx^-dx^\perp (F_{+-}^aF_{+-}^a + F_{12}^a F_{12}^a)/2$.
The ${\cal E}^a$ will enter the definition of the canonical 
variable $E^a = {\cal E}^a +...$ conjugated to $A_-^a$ 
(see below Eq. (\ref{exprE})). The terms
of $E^a$ denoted by the dots will come from  the terms with the time 
derivatives of $A_k^a$ expressed in terms of the new 
variables $\tilde A_k^a$.

The new variables $\tilde{A}_k^a$ of the next stage of the variable 
transformations are connected
with the initial variables $A_k^a$ by a gauge transformation
depending on $A_-^a$:
\begin{equation}
\label{tildedef}
 \tilde A_k^a\frac{\sigma^a}{2}=U^\dagger\bigg(A_k^a\frac{\sigma^a}{2}-
\frac{1}{ig}\partial_k \bigg) U,
\end{equation}
where $\sigma^a$ are the Pauli matrices, and $U$ is a matrix
of $SU(2)$ depending on $A_-^a$. 
The gauge transformation is the one that connects
the initial field configuration to the corresponding configuration
in the light-cone gauge: $\tilde{A}_-^a\sigma^a/2\equiv 
U^\dagger(A_-^a\sigma^a/2-
\partial_-/(ig))U= \sigma^3 \alpha/2$,
where $\alpha$ is a functional of $A_-^a$ independent of 
$x^-$. 
Below we systematically use the tilde over a quantity to denote
the quantity gauge transformed to the light-cone gauge.
The crucial point 
is that the transformation from $A_k^a$ to $\tilde{A}_k^a$ is one-to-one
at fixed $A_-^a$.
Obviously, the transformation from $A_-^a$ to $\tilde{A}_-^a$ is not 
one-to-one. The transformation from $A_+^a$ to $\tilde A_+^a$
is one-to-one, but it involves time derivatives of $A_-^a$.
Thus, we keep the initial configurations $A_\pm^a$
as independent variables and consider the variables $\tilde{A}_\pm^a$
as functionals of the independent variables. For an illuminating discussion
of the gauge transformation to the light-cone gauge see, for example,
Ref. \cite{Lenz}, where explicit formulas for $U$
can be found.

We now need to express the term $F_{-k}^aF_{+k}^a$ of Eq. (\ref{calE}),
which contains the time derivatives of $A_k^a$, in terms of the new set of 
variables ${A_\pm^a, \tilde A_k^a}$. By the gauge invariance,
the form of $F_{-k}^aF_{+k}^a$ in terms of ${\tilde A_\pm^a, \tilde A_k^a}$
is known: it is $\tilde F_{-k}^a\tilde F_{+k}^a$.
Thus, we need to express $\tilde A_\pm^a$ in terms of $A_\pm^a$.
The connection between $A_-^a$ and $\alpha$ (recall that 
$\tilde A_-^a=\delta^{3a} \alpha$) is easy to find considering
a gauge invariant quantity expressible in terms of the component
$A_-^a$ alone. It is the trace of the large Wilson loop embracing
the whole span of the compactified direction $x^-$. Thus, in what follows, 
we will treat $\alpha$ as a known functional of $A_-^a$.

To express $\tilde A_+^a$ in terms of $A_\pm^a$, we need to introduce  
special bases in the space of field configurations. The connection will
be found between the coefficients of the expansions of the tilded
and untilded fields over the tilded and untilded bases. These
expansions are an important ingredient of our approach.

From now on, we switch over from the components with color indices to
matrices: $A_\mu=A_\mu^a \sigma^a/2$, $\sigma^a$ are the Pauli
matrices. We will consider
two sets of bases in the space of field configurations, 
$\chi_p$ and $ \tilde\chi_p$, where $p$ is an index described below,
and every $\chi_p,\,\tilde\chi_p$ is a traceless matrix field depending
on the space-time coordinates.

The bases are complete and orthonormal, 
$\langle\chi_p|\chi_{p'}\rangle = \delta_{pp'},\,
\langle\tilde\chi_p|\tilde\chi_{p'}\rangle = \delta_{pp'}$, 
with respect to the following scalar 
product in the space of matrix fields:
\begin{equation}
\label{scprod}
\langle\Phi|\Psi\rangle = \int\,dx^-\,2 \,{\rm Tr}\, \Phi^\dagger\Psi.
\end{equation}
Therefore, any field is expressible as a sum over the bases:
$A_\mu=\sum_p\chi_p\langle\chi_p|A_\mu\rangle,\,
\tilde A_\mu=\sum_p\tilde\chi_p\langle\tilde\chi_p|\tilde A_\mu\rangle$.
For the components of the fields with respect to the bases, 
we will use the notation
$A_\mu^p=\langle\chi_p|A_\mu\rangle$. The fields will be
expanded only over the corresponding bases: tilded over tilded
basis, untilded over untilded. Note that the expansion coefficients
are independent of $x^-$.

The bases $\chi_p$ and $\tilde\chi_p$ are the bases of eigenfunctions
of the operators $\hat p\equiv D_-/i$ and $\hat{\tilde p}
\equiv\tilde D_-/i$, respectively:
\begin{equation}
\label{basedef}
\hat p \chi_p=p\chi_p,\,
\hat{\tilde p}\tilde\chi_p=p\tilde \chi_p.
\end{equation}
Note that both $\hat p$ and $\hat{\tilde p}$ are Hermitian with
the scalar product (\ref{scprod}). Intuitively, the eigenvalues
$p$ correspond to the values of the quanta of the longitudinal
momentum $P_-$. It is easy to understand that they are gauge invariant.
Thus, there is no tilde over the eigenvalue $p$ in the second
of the Eqs. (\ref{basedef}).
We also indiscriminately interchange the label on the eigenfunction
and the eigenvalue.

Now we give an explicit form of $\tilde\chi_p$. Recalling the 
definitions
$\tilde A_-^a=\delta^{3a} \alpha$ and 
$\tilde D_-\Phi = \partial_-\Phi -ig[\tilde A_-,\Phi]$, 
it is easy to see that the
eigenvalues of $\hat{\tilde p}$ are
\begin{equation}
\label{eigenvalues}
p(n,\sigma) = \frac{2\pi n}{L}+\sigma g\alpha,\, \sigma=-1,0,+1,
\end{equation}
where $n$ is a number of the Fourier mode, and $\sigma$ is an index
to label the splitting of the level due to the presence of the 
gauge field. And the corresponding
eigenfunctions are
\begin{equation}
\label{tildedeigenfunctions}
\tilde\chi_{p(n,0)}=\frac{\exp(i\frac{ 2\pi n}{L}x^-)\sigma^3}{2 \sqrt L},\,
\tilde\chi_{p(n,\pm)}=\frac{\exp(i\frac{ 2\pi n}{L}x^-)
(\sigma^1\mp i\sigma^2)}{2\sqrt{2 L}}.
\end{equation}

Now, we give  a representation for $\chi_p$.
If $U$ is the unitary matrix of the gauge transformation from
$\tilde A_\mu$ to $A_\mu$, $A_\mu=U(\tilde A_\mu-
\partial_\mu/(ig))U^\dagger$, it is easy to check
that 
\begin{equation}
\label{eigenfunctions}
\chi_p=U\tilde\chi_p U^\dagger.
\end{equation}
In words: the eigenfunctions are transformed uniformly under
the gauge transformations. The crucial difference between
the tilded and untilded bases is that the one corresponding to
the light-cone gauge is independent of the time and transverse
space coordinates. We will exploit this property of the tilded basis
in what follows.

Now we are ready to express $\tilde A_+^p$ in terms of
$A_\pm^p$. To this end, consider a gauge invariant
object $\langle\chi_p|F_{-+}\rangle$ at nonzero $p$ (it is gauge invariant,
because both $\chi_p$ and $F_{-+}$ are transformed uniformly and the 
scalar product involves the trace).
Calculated in the light-cone gauge, it is $ip\tilde A_+^p$
(to see this, note that $\langle\tilde\chi_p|\dot{\tilde A}_-\rangle$ 
is vanishing at nonzero $p$). While in terms of the initial variables,
it is $ip A_+^p-\dot A_-^p$. Thus, we have
\begin{equation}
\label{aplus}
\tilde A_+^p =\frac{1}{ip}F_{-+}^p=A_+^p-\frac{1}{ip}\dot A_-^p,\, p\neq 0.
\end{equation}
These expressions are summarized as follows: $\tilde A_+$ is linear in
$A_+$ and $\dot A_-$. This observation helps to keep track of the 
terms of the action involving the time derivatives of
$A_-$.

To express the {\it zero mode} of $\tilde A_+$, 
$\langle\tilde\chi_0|\tilde A_+\rangle\equiv\tilde A_+^0$, in terms of
$A_\pm$, we consider another
gauge invariant object, $\langle\chi_p|D_+\chi_p\rangle$, for an eigenvector
$\chi_p$ with nonzero commutator $[\chi_p,\chi_p^\dagger]=\epsilon(p)\chi_0$. 
The function $\epsilon(p)$ above is gauge invariant and easy
to calculate using Eq. (\ref{tildedeigenfunctions}).
Calculated in the light-cone gauge,  
$\langle\chi_p|D_+\chi_p\rangle=-ig\epsilon(p)\tilde A_+^0$,
while in the initial variables it is 
$\epsilon(p)(\langle\tilde \chi_0|U^\dagger\partial_+U\rangle-ig
A_+^0)$. To obtain the last expression, 
Eq. (\ref{eigenfunctions}) was used. Thus,
\begin{equation}
\label{zeromode}
\tilde A_+^0=A_+^0
-\frac{1}{ig}\langle\tilde\chi_0|U^\dagger\partial_+U\rangle.
\end{equation}

Now we go back to the term 
$\langle \tilde F_{-k}|\tilde F_{+k}\rangle$ of the action (\ref{calE}).
Using Eq. (\ref{aplus}), the representations 
$\tilde F_{-k}=\tilde D_-\tilde A_k-\partial_k\tilde A_-,\,
\tilde F_{+k}=\dot {\tilde A_k}-\tilde D_k\tilde A_+$,
and the completeness of the base $\tilde\chi_p$,
we derive
\begin{eqnarray}
\label{keytransf}
\int\,dx^+dx^\perp \langle \tilde F_{-k}|\tilde F_{+k}\rangle
= \int\,dx^+dx^\perp\Big[ \langle \tilde D_-\tilde A_k|\dot{\tilde A}_k\rangle
  -\langle\partial_k\tilde A_k|\dot{\tilde A}_-\rangle\nonumber\\
  +\sum_{p\neq 0}\langle \frac{1}{\tilde D_-}\tilde D_k\tilde F_{-k}|
                 \tilde\chi_p\rangle    
                 \langle\chi_p|F_{+-}\rangle + 
                  \big(\tilde D_k\tilde F_{-k}\big)^0 \tilde A_+^0\Big].
\end{eqnarray} 
The second term in the rhs of Eq. (\ref{keytransf}) comes from the product
$\langle\partial_k \tilde A_-|\dot{\tilde A}_k\rangle$, and is obtained
via integration by parts. 
The rhs is linear in $F_{+-}$. The coefficient
by $F_{+-}$ will give a contribution to $E$, the canonical variable
conjugated to $A_-$. Another contribution to $E$ will come from the
term $\langle\partial_k\tilde A_k|\dot{\tilde A}_-\rangle$ 
of Eq. (\ref{keytransf}) because $\dot{\tilde A}_-$ is linear in $\dot A_-$.
To express $\dot{\tilde A}_-$ in terms of $\dot A_-$, we use 
Eq. (\ref{eigenvalues})
and the definition $\tilde A_- = \sqrt L \tilde\chi_0\alpha$,
and express the time derivative in  terms of the gauge invariant $p$:
\begin{equation}
\dot{\tilde A_-}=\frac{\sqrt L}{g} \tilde \chi_0\dot p(n,+).
\end{equation}
The time derivative of the gauge invariant eigenvalue
can be calculated as a time derivative of an expectation 
value over the untilded
eigenvector of the untilded operator $\hat p$:
\begin{equation}
\label{dotA}
\dot p(n,+) = \langle \chi_{p(n, +)}|\dot{\hat p}\chi_{p(n, +)}\rangle
            =\frac{g}{\sqrt L}\langle\chi_0|\dot A_-\rangle,
\end{equation}
where we have taken into account that 
$[\chi_{p(n,-)},\chi_{p(n,+)}]=\chi_0/\sqrt L$.

With Eqs. (\ref{keytransf})--(\ref{dotA}), action (\ref{calE})
can be transformed to
\begin{eqnarray}
\label{E}
S_{glue}=\int\,dx^+dx^\perp\Big[\langle E|\dot A_-\rangle+
                   \langle\tilde D_-\tilde A_k|\dot{\tilde A}_k\rangle
-\frac{1}{2}\big(\langle{\cal E}|{\cal E}\rangle + 
                  \langle\tilde F_{12}|\tilde F_{12}\rangle\big)\nonumber\\
-\langle E|D_-A_+\rangle + (\tilde D_k\tilde F_{-k})^0\tilde A_+^0\Big],
\end{eqnarray}
where
\begin{equation}
\label{exprE}
E={\cal E} - \chi_0\partial_k \tilde A_k^0 + \sum_{p\neq 0}
\chi_p \bigg(\frac{1}{\tilde D_-}\tilde D_k\tilde F_{-k}\bigg)^p.
\end{equation}
In the above Eq. (\ref{E}), 
$E$ and $A_-$ are the canonically conjugated variables,
the content of the round brackets is the Hamiltonian, and
the last line contains the Lagrange multiplier $A_+$.
Eq. (\ref{exprE}) is to be used to express ${\cal E}$
of the Hamiltonian in terms of the canonical variables.

The last step in the variable change is to reveal the canonical
variables connected with $\tilde A_k$. Using the completeness of
the basis $\tilde\chi_p$, we rewrite the relevant term:
$\langle \tilde D_-\tilde A_k|\dot{\tilde A}_k\rangle=
i\sum_{p\neq 0}p (\tilde A_k^p)^\dagger\dot{\tilde A_k^p}$, where
the dagger means complex conjugation. Note that the independence
on the tilded basis of the time was crucial: we substituted the
projections of the time derivative $\dot {\tilde A}_k$
on the basis vectors by the
time derivatives of the projections. Notice now that $A^{(-p)}=(A^p)^\dagger$
for any Hermitian  field $A$. This observation makes evident that
\begin{equation}
\label{laststroke}
\langle \tilde D_-\tilde A_k|\dot{\tilde A}_k\rangle= 
i\frac{1}{2}\sum_{p> 0} \big[(a^p_k)^\dagger\dot a^p_k-
                                 (\dot a^p_k)^\dagger a^p_k\big],
\end{equation}
where
\begin{equation}
\label{cv}
a^p_k = \sqrt{2p} \tilde A_k^{(-p)},\, p>0.
\end{equation}
Up to the total time derivative, the rhs of Eq. (\ref{laststroke}) 
takes the canonical form $\sum_{p>0}P^p_k\dot{Q}^p_k$
after the substitution $a^p_k=(Q^p_k+iP^p_k)/\sqrt 2$. We conclude that
$a^p_k(x^\perp)$ are the canonical variables with the following 
Poisson bracket:
\begin{equation}
\label{Pb}
\{a^p_k(x^\perp),(a^q_l)^\dagger(y^\perp)\}=i\delta^{pq}\delta_{kl}
                           \delta(x^\perp-y^\perp).
\end{equation}
Notice that the lhs of Eq. (\ref{laststroke}) expressed with Eq.
(\ref{cv}) in terms of the canonical variables could contain the time
derivatives of $p$, but these terms are cancelled against each other.

An important explanation is in order: Eq. (\ref{laststroke}) contains 
the inequality
$p>0$. It may seem that the fulfillment of this inequality depends
on the configuration of $A_-$. This  though is not the case, because
we can treat the $g\alpha$ of Eq. (\ref{eigenvalues}) 
as the splitting of the levels of $\hat p$. As such, it is constrained
by the inequality
\begin{equation}
\label{constraint}
0\leq\alpha\leq\frac{\pi}{gL}.
\end{equation}
In fact, the whole construction can be reformulated without
use of the transformation to the light-cone gauge. In this 
reformulation,
$g\alpha$ is {\it defined} as the minimal splitting between the 
levels of $\hat p$, and $\chi_p$ are {\it defined} as the eigenfunctions 
of $\hat p$. After this explanation, we can label the canonical
variables $a^p_k$ of Eq. (\ref{cv}) by $n>0$ and the discrete variable
$\sigma=-1, 0,+1$ related to $p$: $p=2\pi n/L+\sigma g\alpha$. 
There are also  degrees of freedom $a^p_k$ corresponding to
$n=0$, $p=g\alpha$. We will stick in what follows to the more compact
labeling by the eigenvalue $p>0$.

It is important that the dynamical variables $a^p_k$ are not in 
one-to-one correspondence with $\tilde A_k$. The latter contains more information.
Namely, it contains a {\it zero mode} $B_k$($\equiv \tilde A^0_k$):
\begin{equation}
\label{zm}
\tilde A_k=B_k\tilde\chi_0 + \sum_{p>0}\bigg[
          \frac{\tilde\chi_p}{\sqrt{2p}}(a^p_k)^\dagger+
           \frac{\tilde\chi_p^\dagger}{\sqrt{2p}}a^p_k\bigg].
\end{equation}

The crucial observation is that the zero mode $B_k$ enters only the 
Hamiltonian
in the round bracket of Eq. (\ref{E}), and not the terms involving
the Lagrange multiplier $A_+$. The variation with respect to
the latter gives the light-front version of the Gauss law.
It consists of two components. The first component comes from the
variation of the first term of the second line of Eq. (\ref{E}), 
while the second component comes from the second term. 
Note that these terms
can be varied independently, since the first term contains
only the nonzero modes of $A_+$, while the second term depends only 
on the zero
mode $\tilde A_+^0$. The variation of the Hamiltonian with
respect to $B_k$ gives an equation linear in $B_k$. Solving it,
one determines $B_k$ in terms of the canonical variables.

\section{The Gauss Law}

Next, we write down the components of the Gauss law. The first component
is
\begin{equation}
\label{glf}
D_-E=0,
\end{equation}
and the second,
\begin{equation}
\label{gls}
\Delta_\perp \tilde A_-^0 + g 
\sum_{p>0}\epsilon(p)(a^p_k)^\dagger a^p_k=0,
\end{equation}
where $\Delta_\perp\equiv\partial_1^2+\partial_2^2$, and 
$\epsilon(p)$ was previously introduced by the relation
$[\chi_p,\chi_p^\dagger]=\epsilon(p)\chi_0$. Note that the
first component holds at any space-time point, while the
second is independent of $x^-$ and holds at any point of the transverse
plane at any moment of the light-front time. Also note that
$\epsilon(p)=-\sigma/\sqrt L$, where
$p=2\pi n/L+\sigma g\alpha$.

The natural next step is to calculate the Poisson brackets
between the canonical variables and the components of the 
Gauss law smeared with some weights. The expectation is
that the components of the Gauss law will generate the
local gauge transformations of the canonical variables.
A simple calculation supports this expectation. The
first component of the Gauss law (\ref{glf}) generates
the gauge transformation of the canonical variables
with the local parameter $\phi$ orthogonal to the zero mode,
$\langle\chi_0|\phi\rangle=0$. As the form of Eq. (\ref{glf})
shows, this transformation leaves $a^p_k$ invariant. 
That we already know from the definition (\ref{cv}).
To see this, notice that $\tilde A_k^p=F_{-k}^p/(ip)$
at nonzero $p$. The second component of the Gauss law
(\ref{gls}) generates the leftover Abelian gauge transformations
whose local parameters are proportional to the zero mode,
$\phi=\psi\chi_0$, where $\psi$ is a (non-matrix) field 
independent of $x^-$. As the form of Eq. (\ref{gls}) shows,
the $a^p_k$ are transformed under these transformations as
charged fields under an Abelian gauge transformation,
and the charge is determined by the sign $\sigma$. This 
can also be understood independently from the definition (\ref{cv}).
To see this, notice that the Abelian transformations leave
$A_-$ invariant. If we take the basis $\chi_p$ to be completely 
determined by the configuration of $A_-$, the basis $\chi_p$
is also invariant under these transformations. It follows from this 
observation that $F_{-k}^p$ transforms as a charged field under
the Abelian gauge transformations. We conclude that
the components of the light-front Gauss law generate the
gauge transformations, which is quite analogous to the situation
in the conventional equal time formulation. 

Note how the zero mode component of the Gauss law (\ref{gls}) 
acts on $E$: 
$\{\int\psi\Delta_\perp \tilde  A^0_- ,E\}=-\Delta_\perp\psi\chi_0$.
This is again in accord with our expectations, because it corresponds
to the invariance of the zero mode of ${\cal E}=F_{+-}$ under the Abelian
subgroup of the gauge transformations (see Eq. (\ref{exprE}) for the
connection between ${\cal E}$ and $E$). To understand it, assume that
$B_k\equiv\tilde A_k^0$ transforms as it should under the gauge 
transformation generated by $\psi\chi_0$. Then, there is a cancellation
between the transformation of $E$ and the transformation
of $B_k$ in Eq. (\ref{exprE}) which leaves the zero mode of ${\cal E}$
invariant as promised.

To finally check the assumption that the components of the Gauss
law generate the gauge transformation of all the fields involved, 
we should find $B_k$ in terms of the
canonical variables and check that the Poisson brackets between them
and the components of the Gauss law generate the gauge transformations
of $B_k$. We postpone this check to make  the following observations.
As the components of the Gauss law generate the gauge transformations
of the canonical variables, the Poisson brackets of these components
form a closed algebra and commute with the Hamiltonian, because the
latter is gauge invariant. Therefore, the light-front gluodynamics
is a generalized dynamical system with  first class constraints.
This is the main result of the paper. Another observation is that
the canonical light-front formulation leads to a natural Abeliazation
of the theory. All the non-Abelian transformations act
nontrivially only on $A_-$, leaving the rest of the dynamical variables
invariant. An Abelian subgroup of the gauge  transformations 
which leaves $A_-$
invariant is naturally singled out. The zero mode of the gauge
field plays the role of the Abelian gauge field in an Abelian 
gauge theory
with the space-time dimension decreased by one (the $x^-$ dimension
is ``eaten'' by the projection on the zero mode). This Abelian 
gauge field turns out to be nondynamical and expressible in terms of
the dynamical variables.

\section{The Zero Mode Equation}

We now work out the equation for $B_k$. To this end, we take a variation
of the round bracket of Eq. (\ref{E}) over $\tilde A_k$, convolute it with
$\tilde\chi_0$, and require the convolution to vanish. It gives 
\begin{equation}
\label{B}
\bigg(-\Delta_\perp  + \frac{g^2}{L}\sum_{p\neq 0,l}|\tilde A^p_l|^2\bigg)B_k=
\partial_kE^0+ig\sum_{p\neq 0}\epsilon(p)\bigg(
E-\frac{1}{\tilde D_-}\tilde D_l\tilde F_{-l}\bigg)^p(\tilde A^{p}_k)^\dagger-
\epsilon_{kl}\big(\tilde D_l\tilde F_{12}\big)^0,
\end{equation}
where $B_k$ is set to zero 
in the rhs, and
$\epsilon_{kl}$ is the antisymmetric tensor with $\epsilon_{12}=1$.
This is the equation to determine $B_k$. First we note that it gives
$B_k$ which transforms in the right way under the gauge transformations
of the rhs. Second, to solve for $B_k$, we need to invert the operator
acting on $B_k$. It is a Schr\"odinger operator with a positive potential.
Thus, it may have a zero eigenvalue only at $a^p_k=0$. In this case,
the potential vanishes, and the equation does not restrict the contribution
to $B_k$ which does not depend on $x^\perp$. Though, in this latter case,
$B_k$ enters the Hamiltonian only via its derivatives over the transverse 
coordinates. Thus, we conclude that Eq. (\ref{B}) suffices to determine
the Hamiltonian of the light-front gluodynamics.

We mention here that Eq. (\ref{B}) is in agreement with the equation 
obtained for the zero modes in Ref. \cite{Nov}. In Ref. 
\cite{Nov}, the equation 
for the zero modes was obtained after gauge fixing. We stress that there is
no need to fix the gauge to derive Eq. (\ref{B}). The operator acting on $B$
in the lhs of Eq. (\ref{B}) is gauge invariant.

\section{Quantization and Perturbation Theory}

It is possible to write down the expression for the Hamiltonian with $B_k$
excluded. This completes the formulation of the classical theory. The 
quantization consists in replacing the canonical brackets with 
commutators, and
in prescribing the ordering of the operators in the Hamiltonian. If we 
consider the classical Hamiltonian with the Gauss law implemented, 
there is no ambiguity in ordering the operators $E$ and $A_-$, because they 
do not join one another in the Hamiltonian. But there is certainly     
an ambiguity of ordering for the transverse gluons creation-annihilation
operators. We choose the simplest prescription of 
normal ordering. 

Making this choice, we may contradict the results
of the equal-time quantization, where its own ordering is assumed.
That is, equal-time normal ordering
may not coincide with the light-front normal ordering we have chosen.
Thus, we must confirm that our prescription does not lead to a contradiction.

Here we demonstrate that there is no contradiction between the normal ordering
light-front quantization and the standard perturbation theory in covariant
gauges. 

In checking this, we benefit from the fact that 
the obtained formulation is standard, 
and we can use the methods of Ref. \cite{Faddeev} to write down the functional
integral representation for vacuum expectations of the time ordered products of
gauge invariant local operators 
(for example, for the products of $F_{\mu\nu}^2$). The time ordering is
understood with respect to the light-front time $x^+$. In accord 
with that standard methods, the functional integral will run over
the canonical variables, and the additional integration over $A_+, 
\tilde A^0_+$ will impose the Gauss law on the states featuring the derivation
of the functional integral representation. To simplify the treatment we do not 
impose any boundary conditions, and consider instead a finite span $2T$ of the 
light-front time $x^+$ for the fields in the functional integral. 
When, in the limit $T\rightarrow\infty(1-i\epsilon)$, 
the vacuum expectation is reproduced (up to normalization) due to the presence
of $i\epsilon$ in the time span.

Next, we derive the Hamiltonian in the reverse route: 
introduce $B_k$ in the functional integral and restore
it back in the action, go over from the 
integration over $a^\dagger, a, B_k$ to the integration over $A_k$ (see Eq.
(\ref{zm})), shift $E$ to $\cal E$ (see Eq. (\ref{exprE}), 
and, finally, integrate $\cal E$ out. This 
gives us the standard representation in terms of the functional integral over
$A_\mu$, with the standard action $S_{glue}$ in the exponential. However, 
there is a subtle point: on the way back we pick up some determinants. They 
will potentially cause a difference between our formulation and the 
standard equal-time formulation. We now show that this difference 
vanishes in the limit of infinite 
volume, if dimensional regularization is used to regularize the infinite
volume theory.

We pick up the first determinant when the integration over $B_k$ is introduced.
$B_k$ enters the Hamiltonian quadratically. So, 
integrating out $B_k$ will return
the initial expression for the Hamiltonian with $B_k$ excluded via 
Eq. (\ref{B}), and also will produce a power of the determinant of the 
operator featuring the lhs of Eq. (\ref{B}), ${\rm det}(-\Delta_\perp +
g^2\sum_p |\tilde A^p|^2/L)$. The dependence of this determinant on the fields is
explicitly suppressed by the inverse $L$. Therefore, in the limit of
infinite $L$, this determinant introduces only an (infinite) multiplication 
constant in the measure of integration.

The second determinant appears when we go over from the integration
over $B_k, a^\dagger, a$ to the integration over $A_k$. It is related to the
presence of the denominators $\sqrt{2p}$ in Eq. (\ref{B}). 
Since $p$ are the eigenvalues of
$D_-$, the new determinant is a power of the determinant of $D_-$.
And it is known that ${\rm det}D_-$ is a constant
independent of $A_-$, if the dimensional regularization is used to 
regularize the infrared divergences (see, for example, Ref. \cite{Bassetto}).

We conclude that no determinants appear in the limit of infinite volume when
the dimensional regularization is used.

The next step is to fix the gauge in the functional integral. Here we again 
enjoy our independence of the gauge fixing:  we are not forced to consider the 
light-cone gauge, which has extra singularities in the gluon propagator.
Instead, we consider covariant gauges, introducing Faddeev-Popov ghosts in 
the standard way. 

As a result, we obtain the standard representation for the Green's functions
in terms of the functional integral. The only trace of the light-front
approach left is the restriction of the fields to the strip 
$-T \leq x^+\leq T$. As we do not restrict the 
boundary values at $|x^+| = T$ of the fields under the functional 
integration in any way, the classical 
solution of the linearized equations of motion related to the determination of
the propagators should vanish on the boundaries of the strip. It is easy
to check that the imaginary shift in $T$ allows the presence of the 
unique solution to the classical linearized field equations which 
boundary values
vanish at $T\rightarrow \infty(1-i\epsilon)$. Then, it is easy to check that
the propagators appearing have the Feynman $i\epsilon$ in the denominators.

We conclude that the normal ordering light-front quantization reproduces the
standard Feynman rules in covariant gauges. We stress that the above derivation
is possible only because the definition of the canonical variables we found
does not depend on any gauge fixing.

\section{The Non-Abelian Plane Waves Reduction}

Now we turn to the analysis of the reduced dynamics by requiring 
$P_-$ to vanish. To see the restriction this requirement sets on the
dynamical variables,
consider $P_-$ as it is yielded by the Noether procedure:
\begin{equation}
\label{P-}
P_-=\int\,dx^\perp \langle F_{-k}|F_{-k}\rangle.
\end{equation}
In terms of the canonical variables, it is
\begin{equation}
\label{Pcv}
P_-=\int\,dx^\perp \big[\sum_{p>0}p(a^p_k)^\dagger a^p_k-
\tilde A_-^0\Delta_\perp\tilde A_-^0\big].
\end{equation}
With the zero mode component of the Gauss law (\ref{gls}) 
and the relation $\tilde A_-=\sqrt L\alpha\tilde\chi_0$ taken into account,
it is
\begin{equation}
\label{Pfinal}
P_-=\int\,dx^\perp \sum_{p>0}\frac{2\pi n}{L}(a^p_k)^\dagger a^p_k,
\end{equation}
where $n$ is related to $p$ by $p=2\pi n/L+\sigma g\alpha$.
Thus, we see that vanishing of $a^p_k$ with $p>g\alpha$
is necessary and sufficient for vanishing of
$P_-$. We note in passing that $P_-$ is independent of the canonical
pair $E$, $A_-$. $A_-$ enters $P_-$ only indirectly 
via the
zero mode component of the Gauss law (\ref{gls}). 
The latter restricts the possible configurations of $a^p_k$.
In particular, integrating Eq.
(\ref{gls}) over the transverse plane we see that the total sum of
the Abelian charges over the transverse plane should vanish all the time.
This, in fact, implies that the only $a^p_k$ which is 
allowed by 
the vanishing of $P_-$ ($p=g\alpha$) is forbidden by the Gauss law,
because it has definite Abelian charge, and nothing can compensate it. 
Thus, we conclude that
the reduction to the configurations of vanishing $P_-$ is equivalent to
the reduction to vanishing $a^p_k$, {\it i.e.}, to the reduction from 
Eq. (\ref{zm}) to $\tilde A_k = B_k\tilde\chi_0$.

This reduction simplifies the action (\ref{E}) dramatically. The second term
of the square bracket vanishes, the round bracket is simplified because
${\cal E}$ becomes just $E+\chi_0\partial_kB_k$ (see Eq. 
(\ref{exprE}) and 
recall that $B_k\equiv\tilde A^0_k$), $\tilde F_{12}$
becomes $(\partial_1 B_2-\partial_2 B_1)\tilde\chi_0$, and the last term 
of the square bracket becomes
$(\partial_k \tilde A_-^0\partial_k\tilde A_+^0)$. 
At no compactification over the $x^-$ direction, in the light-cone
gauge $\tilde A_-=0$, and at boundary conditions
on $E$ at the $x^-$ infinity suppressing the zero mode $E^0$, 
the classical equations of the reduced dynamics
implied by this action are $\partial_-\tilde A_+=0$, 
$\Delta_\perp \tilde A_+=0$, $\tilde A_k=0$. 
This reduction
of the classical gauge equations is known since the work of
Coleman \cite{Col}. Thus, we may say that the reduced dynamics
of the paper
generalizes  one of Coleman's non-Abelian plane waves for
the case of the compactified $x^-$. Our next aim is to
see the Hamiltonian formulation of the paper at work. We will
find the quantum spectrum of the non-Abelian plane waves.

To this end, we will work in the gauge $D_-A_+=0$, $\tilde A_+^0=0$,
which is the light-front analog of the Weyl gauge. In this gauge, the
reduced system is a Hamiltonian system with the Hamiltonian
\begin{equation}
\label{HredB}
H_{red}(B)=\int\, dx^\perp \frac{1}{2}\Big[\big(E^0+\partial_k B_k\big)^2
+\big(\partial_1B_2-\partial_2 B_1\big)^2\Big].
\end{equation}
Minimization with respect to $B_k$ gives
\begin{equation}
\label{Hred}
H_{red}=\frac{1}{2}\bar E^2 V_\perp,
\end{equation}
where
\begin{equation}
\label{barE}
\bar E\equiv\int\frac{dx^\perp}{V_\perp}\langle\chi_0|E\rangle,
\end{equation}
and $V_\perp\equiv\int dx^\perp$ is the volume of the transverse plane.
Apart from the Hamiltonian (\ref{Hred}), the reduced system is defined
by the Gauss law $D_-E=0$, $\Delta_\perp\tilde A_-^0=0$. Notice that the 
Hamiltonian vanishes in the limit of the infinite transverse volume
on the configurations of $E$ with finite $\int dx^\perp E^0$, because
$\bar E\sim 1/V_\perp$ in this case.

Let us demonstrate that the phase space of the reduced system is
parameterized by two variables. One is $\bar E$, another is 
$\bar A\equiv\int\,dx^\perp\tilde A_-^0/V_\perp$.
This is so because of all $E^p$ only $E^0$ is nonzero by the
component of the Gauss law $D_-E=0$, and, under a gauge transformation,
$E^0$ can be shifted by
the transverse Laplacian of a gauge parameter field (recall that
$\{\int\psi\Delta_\perp \tilde  A^0_- ,E\}=-\Delta_\perp\psi\chi_0$).
Thus, the only piece of $E$ which is simultaneously gauge invariant
and allowed by the Gauss law is $\bar E$. The same holds with respect
to $A_-$: the only gauge invariant component of $A_-$ is 
$\tilde A_-^0$,
and it should be independent of the transverse coordinates by
the zero mode of the Gauss law. 
The reduction to a single gauge invariant component is achievable
because $A_-$ can be transformed to $\tilde A_-^0$ by 
a gauge transfromation, and the latter is a gauge invariant
expressible in terms of the gauge invariant trace of a Wilson loop.

Notice the specific duality between
$E$ and $A_-$: The Gauss law forbids nonzero $E^p$ at $p\neq 0$, while 
the gauge invariance admits only constant $E^0$; for
$A_-$, the Gauss law and the gauge invariance interchange
their roles in the reduction, {\it i.e.}, the gauge invariance forbids
nonzero $\tilde A_-^p$ at nonzero $p$, while the Gauss law 
admits only
constant contributions to $\tilde A_-^0$. In this reasoning, 
we assumed that
the theory
is compactified also in the transverse directions $x^{1,2}$.

Thus, we introduce two variables
$Q\equiv\bar A_-$, and $P\equiv V_\perp\bar E$. They suffice to 
parameterize the phase space of the reduced system, and the normalization
is chosen to make them canonically conjugated: $\{P, Q\}=1$.
The Hamiltonian (\ref{Hred}) in terms of these variables is
$H_{red}=\bar P^2/(2V_\perp)$. Notice that $V_\perp$ plays the role of
the mass of a free nonrelativistic particle. The next
crucial point is that
$Q$, in fact, should be compactified.

To see this, consider a ``large'' gauge transformation generated by the
Hermitian traceless matrix 
$\omega\equiv gx^-\big(2\pi/g\sqrt L\big)\chi_0$. The transformed
gauge field $A_-^U=U(A_--\partial_-/(ig))U^\dagger$, $U=\exp(i\omega)$,
is periodic in $x^-$, if $A_-$ is. Thus, $U$ is indeed a gauge 
transformation of the compactified theory. 
Notice that this is not the case 
for any $\omega'=\lambda\omega$, where $\lambda$ is noninteger. Because 
of this, the above transformation cannot be continuously transformed to
the trivial transformation. 
The presence of the ``large'' gauge transformations in a compactified
theory is known since the work \cite{tHooft}.
In the context of the light-front formulation, 
``large'' gauge transformations have also been considered
in Ref. \cite{Pauli}, and  utilized 
in Ref. \cite{Martinovic} for the Schwinger model.
The particular ``large''
gauge transformation we are considering here is a light-front analog
of the equal-time finite-volume ``central conjugations'' of Ref.
\cite{Luscher}. 
It is easy to check that this transformation
leaves $P$ invariant, and shifts $Q$: $Q\rightarrow Q + 2\pi/g\sqrt L$.
Thus, the theory should be invariant with respect to this shift, because
it is a remnant gauge transformation. The invariance is achieved by the
condition on the wavefunctions in the $Q$-representation:
\begin{equation}
\label{theta}
\Psi\bigg(Q+ \frac{2\pi}{g\sqrt L}\bigg)=e^{i\theta}\Psi(Q),
\end{equation}
where $\theta$ is an angle, $0\leq\theta<2\pi$, parameterizing the
theory.

In fact, the allowed values for the $\theta$ in Eq. (\ref{theta})
are $0$ and $\pi$. This is the case because the double action of the above
large gauge transformation on the gauge fields leaves both $P$ and $Q$
invariant. To see this, notice that $Q$ is expressible in terms of the 
trace of a large Wilson loop, and the latter changes its sign under
the single action of $U$. For more explanations, see 
Refs. \cite{tHooft,Luscher}.
In what follows, we introduce the label $e$ for the superselection
sectors of the theory: $e=0$ for $\theta=0$ and $e=1$ for $\theta=\pi$.
The notation is in accord with the one of 
Refs. \cite{tHooft,Luscher} and is to
remind of the connection with the values of the electric flux.

Then, in the sector $e=0$ the wave functions are periodic,
and in the sector $e=1$, antiperiodic: 
\begin{equation}
\label{sectors}
\Psi(0)=(-1)^e\Psi\bigg(\frac{2\pi}{g\sqrt L}\bigg).
\end{equation}

Condition (\ref{sectors}) singles out a discrete spectrum of the admissible
values for $P$:
\begin{equation}
\label{pn}
P(n)=g\sqrt L\bigg(n+\frac{e}{2}\bigg),
\end{equation}
where $n$ is an integer, and $e$ is either zero or 
unit ``electric flux''.
Recalling that $H_{red}=P^2/(2V_\perp)=P_{+|_{(P_-=0)}}$, we conclude that the
spectrum of $P_+$ in the subspace $P_-=0$ is
\begin{equation}
\label{fin}
P_{+|_{(P_-=0)}}(n)=\frac{g^2 L}{2V_\perp}\bigg(n+\frac{e}{2}\bigg)^2.
\end{equation}
At $n = 0$ and  $e =1$, it coincides with the
``free energy of an electric flux'' of 't Hooft, see
Eq. (9.2) of Ref. \cite{tHooft}.

\section{Discussion and Conclusion}

Our central result, Eq. (\ref{fin}), shows that the presence
of a mass gap in infinite volume theory depends on the ordering 
of the limiting procedure. If one takes first the limit 
$V_\perp \rightarrow \infty$
then there is no  mass gap, and if one takes first $L \rightarrow \infty$
then there is a mass gap. 

We  
note that the dependence of the thermodynamic state on the limiting 
procedure is also
present in  statistical mechanics, where a non-unique limit is 
generally associated with some sort
of first-order phase transition and may indeed be considered
as a possible definition of a phase transition 
\cite{Grif}. On a physically motivated way to select
the ``right'' state see Ref. \cite{Bog}. 

Thus, we interpret the nonexistence of the infinite volume limit
in our case as the indication of the presence of the first-order
phase transition in gluodynamics. This is an acceptable
feature in view of the expected presence of the deconfinement
phase transition and quark-gluon plasma in QCD \cite{plasma}.
Obviously, significant work is needed to reconcile our approach
with what is known about the phase structure of QCD.

Having arrived at such simple spectrum, Eq. (\ref{fin}),
could explain how
previous approximate treatments yielded results seemingly valid
beyond limitations of underlying assumptions \cite{Naus}.

There is similarity of our results with recent works on Yang-Mills fields 
decomposition \cite{Faddeev99,Shabanov} which leads to 
the Abelian dominance \cite{'t Hooft}. However, those equal time
approaches are linked with choice of a proper gauge \cite{Faddeev99} 
or with a nonlocal variable transformation
\cite{Shabanov}, while we have obtained our results without a gauge fixing
in finite volume light-front formulation.

The next steps for development are the generalization
to the $SU(N)$ case and the inclusion of fermions.
In general, the results for the theory with fermions can differ
drastically from pure gauge theory such as the gluodynamics
considered here.

Note that the finite volume light-front formulation 
may play an important
role for string theory,
where one has to quantize a compactified theory.
Since we have obtained a light-front formulation without a gauge fixing 
and in finite volume our results can stimulate a deeper understanding 
of a relation with novel
M-theory developments \cite{Susskind}.

To conclude, we gave a canonical formulation of the 
light-front $SU(2)$ gluodynamics without a gauge fixing. 
The Gauss law was determined and
the system was qualified as a generalized dynamical system with 
first class constraints. 
The formulation obtained reproduces the standard Feynman rules in the
covariant gauges.
The spectrum (\ref{fin}) of the light-front 
Hamiltonian
$P_+$ was determined in the subspace of zero $P_-$ for the case of 
the theory compactified on a torus. An unexpected feature of the
spectrum is that the distance between the levels of $P_+$ may vanish 
in the limit 
of infinite volume, depending on the way the limit is taken. 
This suggests the possibility of the
presence of massless states in certain ``phases'' of the infinite volume 
theory obtained via the limiting procedures with the vanishing 
quantum of $P_+$. There are obvious possibilities to further
develop the formalism: to generalize for $SU(N)$, to include fermions,
to develop the perturbation theory in various gauges at finite and
at infinite volume, {\it etc.} Some of these problems will be addressed in
Ref. \cite{next}.

\acknowledgements The authors are grateful to I. Ya. Aref'eva, 
V. A. Franke, V. A. Kuzmin, \v{L}. Martinovi\v{c}, Yu. V. Novozhilov, 
 A. A. Ovchinnikov, H. Pirner, E. V. Prokhvatilov,
V. A. Rubakov and S. V. Troitskii
for stimulating discussions. They also thank R. Jackiw for communication and 
for pointing out Ref. \cite{Tomb}.
VTK and GBP thank the Fermilab Theory Group for their warm 
hospitality and support.
This work was supported  in parts by the Russian Foundation for Basic
Research, grant No. 00-02-17432, by the NATO Science Programme, 
grant No. PST.CLG.976521, by NSF, grant No. 9970778, and by 
the U. S. Department of Energy, grant No. DE-FG02-87ER40371.

\end{document}